\documentclass[prd,showpacs,amssymb,nofootinbib,twocolumn]{revtex4}
\begin{document}

\title{Chromomagnetic Instability and Induced Magnetic Field in Neutral Two-Flavor Color Superconductivity}
\author{Efrain J. Ferrer}
\author{Vivian de la Incera}
\affiliation{Department of Physics, Western Illinois University,
Macomb, IL 61455, USA}

\begin{abstract}

We find that the chromomagnetic instability existing in neutral two-
flavor color superconductivity at moderate densities is removed by
the formation of an inhomogeneous condensate of charged gluons and
the corresponding induction of a magnetic field. It is shown that
this inhomogeneous ground state is energetically favored over a
homogeneous one. The spontaneous induction of a magnetic field in a
color superconductor at moderate densities can be of interest for
the astrophysics of compact stellar objects exhibiting strong
magnetic fields as magnetars.

\pacs{12.38.Aw, 12.38.-t, 24.85.+p}
\end{abstract}
\maketitle

\section{Introduction}
It is quite plausible that color superconductivity (CS) is realized
in the cold and dense core of compact stellar objects. For matter
densities high enough to guarantee the quark deconfinement, and
sufficiently low to consider the decoupling of the strange quark
mass $M_s$, the quarks can condense to form a two-flavor color
superconductor (2SC) \cite{Bailin-Love}. At higher densities, for
which $M_s$ cannot be regarded as extremely large, the three-flavor
phases $gCFL$ \cite{Alford} and $CFL$ \cite{alf-raj-wil-99/537} will
appear respectively with increasing densities. Matter inside compact
stars should be neutral and remain in $\beta$ equilibrium. When
these conditions along with $M_s$ are taken into account in the
dynamics of the gluons in the $gCFL$ phase (or for that matter in
the 2SC phase if the s-quark is decoupled), some gluon modes become
tachyonic \cite{Igor,Fukushima}, indicating that the system ground
state should be restructured.

The quest to find the stable ground state at the realistic, moderate
densities of compact stars has led to several interesting proposals.
Some of the most promising possibilities are a modified $CFL$-phase
with a condensate of kaons \cite{Schafer} that requires a
space-dependent condensate to remove the instability; a LOFF phase
on which the quarks pair with nonzero total momentum \cite{LOFF};
and a homogenous gluon condensate phase \cite{miransky}. At present
it is not clear if any of these proposals is the final solution to
the problem.

In this paper we will show that there is one more possible solution:
an inhomogeneous condensate of charged gluons that induces a rotated
magnetic field. Our solution is found in the Meissner unstable
region of the so-called gapped $2SC$ phase \cite{Huang}, but we
expect that it could also be realized in the three-flavor theory. It
has the peculiarity of preserving the electromagnetic gauge
invariance $\widetilde{U}_{em}(1)$ of the color superconductor. As
will be shown in our derivations, this inhomogeneous phase is
energetically favored over the one with homogeneous gluon condensate
and no induced magnetic field. We will also discuss its possible
relation to a LOFF-type phase.

In a previous paper \cite{Vortices}, we investigated the formation
of an inhomogeneous gluon condensate in a color superconductor.
However, apart from the fact that in the current work we are going
to consider a two-flavor theory, while in Ref. \cite{Vortices} we
studied a three-flavor system, there is a fundamental difference
between the genesis of the condensation phenomenon that we will
discuss in the present paper and the one found in \cite{Vortices}.
In Ref. \cite{Vortices} the condensation of gluons was the response
of the color superconductor to an externally applied and
sufficiently strong magnetic field. In that case, if no magnetic
field is present, no gluon condensate is formed. Therefore, the
gluon condensation in \cite{Vortices} was a
\textit{magnetic-field-induced phenomenon}. On the other hand, in
this work, no external magnetic field is present to begin with. The
condensation phenomenon here is connected to a Meissner instability
triggered at moderate densities by a pairing stress associated with
the neutrality and $\beta$-equilibrium constraints. Such instability
occurs in the gapped 2SC color superconductor at moderate densities
when the effective Meissner mass of the charged gluons becomes
tachyonic. As we shall show below, in this situation an
inhomogeneous condensate of charged gluons emerges to remove the
chromomagnetic instability created by the pairing mismatch.

In the case of the gluon condensation produced by a sufficiently
strong magnetic field \cite{Vortices}, the anomalous magnetic moment
interaction of the spin-1 charged fields with the external magnetic
field led to a phenomenon of anti-screening, so that the back
reaction of the condensate on the applied field was such that it
boosted the in-medium magnetic field strength. This anti-screening
is a well-known phenomenon that occurs in all the instances of
charged spin-1 field condensation in the presence of a strong
magnetic field \cite{zero-mode, Olesen}. Amazingly, as we will show
in this paper, the system predisposition to boost an in-medium
magnetic field through the magnetic moment of the spin-1 charged
condensate works in favor of the \textit{spontaneous generation} of
a rotated magnetic field in the unstable region of the gapped 2SC
when no external magnetic field is present. Indeed, in this case the
currents associated to the inhomogeneous condensate of charged
gluons induce a non-zero rotated magnetic field in the system.
Hence, the magnetic field here is a consequence of the gluon
condensation, while in \cite{Vortices} it was the other way around.
The spontaneous induction of a rotated magnetic field in CS systems
that have pairings with mismatched Fermi surfaces is a new kind of
phenomenon that can serve to generate magnetic fields in stellar
compact objects as magnetars.

The plan of the paper is as follows. In Sec.\ref{Neutral Two-Flavor
Dense QCD} we present the effective theory of the gluonic phase with
electromagnetic interactions in the neutral two-flavor dense QCD.
The arising of an inhomogeneous gluon condensate to remove the
chromomagnetic instability caused by the Fermi-surface pairing
mismatch in the 2SC color superconductor at moderate densities, as
well as the corresponding induction of a rotated magnetic field are
investigated in Sec.\ref {Inhomogeneous gluon condensate}. Finally,
in Sec. \ref{Astrophysical Connotations} we discuss the significance
of the reported findings for the astrophysics of highly magnetized
compact objects.

\section{Neutral Two-Flavor Dense QCD with Electromagnetic Interactions}\label{Neutral Two-Flavor Dense QCD}

In the gapped $2SC$ phase the solution of the neutrality conditions
$\partial \Omega_{0}/\partial \mu_{i}=0$, with $\mu_{i}=\mu_{e},
\mu_{8}, \mu_{3}$, and gap equation $\partial \Omega_{0}/\partial
\Delta=0$, for the effective potential $\Omega_{0}$ in the
mean-field approximation, led to $\mu_{3}=0$, and nonzero values of
$\mu_{e}$, and $\mu_{8}$, satisfying $\mu_{8}\ll \mu_{e}<\mu$ for a
wide range of parameters \cite{Huang}. Here $\mu$ is the quark
chemical potential, $\mu_{e}$ the electric chemical potential, and
the "chemical potentials" $\mu_{8}$ and $\mu_{3}$ are strictly
speaking condensates of the time components of gauge fields,
$\mu_{8}= (\sqrt{3}g/2)\langle G_{0}^{(8)}\rangle$ and $\mu_{3}=
(g/2)\langle G_{0}^{(3)}\rangle$. The nonzero values of the chemical
potentials produce a mismatch between the Fermi spheres of the quark
Cooper pairs, $\delta \mu=\mu_{e}/2$.

The gapped 2SC turned out to be unstable once the gauge fields
$\{G_{\mu}^{(1)}, G_{\mu}^{(2)}, G_{\mu}^{(3)}, K_{\mu},
K_{\mu}^{\dag}, \widetilde{G}^{8}_{\mu}, \widetilde{A}_{\mu}\}$ were
taken into consideration. As shown in Ref. \cite{Igor}, the gluons
$G_{\mu}^{(1,2,3)}$ are massless, the in-medium $8^{th}$-gluon
$\widetilde{G}_{\mu}^8=\sin{\theta}A_{\mu}+\cos{\theta}\,G^{8}_{\mu}$
has positive Meissner square mass, and the $K$-gluon doublet
$K^{\intercal}_{\mu}\equiv
\frac{1}{\sqrt{2}}(G_{\mu}^{(4)}-iG_{\mu}^{(5)},G_{\mu}^{(6)}-iG_{\mu}^{(7)})$
has Meissner square mass that becomes imaginary for $\Delta > \delta
\mu > \Delta/\sqrt{2}$, signalizing the onset of an unstable ground
state. The mass of the in-medium (rotated) electromagnetic field
$\widetilde{A}_{\mu}=\cos{\theta}\,A_{\mu}-\sin{\theta}\,G^{8}_{\mu}$
is zero, which is consistent with the remaining unbroken
$\widetilde{U}(1)_{em}$ group. Note that even though the gluons are
neutral with respect to the regular electric charge, in the 2SC
phase the gluon complex doublet $K_{\mu}$ has nonzero rotated charge
$\widetilde{Q}=Q-\frac{1}{\sqrt{3}}T_{8}$ of magnitude
$\widetilde{q}=\widetilde{e}/2$, with $\widetilde{e} = e
\cos{\theta}$, $T_{8}$ the $8th$ color generator, and $Q$ the
conventional electric charge operator \cite{Cristina}.

In what follows, we will find a stable ground state solution near
the critical point $\delta\mu_{c}=\Delta/\sqrt{2}$. Close to that
point the absolute value of the square magnetic mass $|m_{M}^{2}|$
is very small, thus allowing analytical calculations. The tachyonic
modes in the Meissner-unstable gapped 2SC are associated only to
charged gluon fields. It is natural to expect then that a new,
stable ground state should incorporate the condensation of these
charged gluons. In general we should allow for an inhomogeneous
condensate. Given that this kind of solution may generate rotated
electromagnetic currents, the rotated electromagnetism should be
included in the general framework of the condensation phenomenon.
With this aim, we start from the mean-field effective action of the
$2SC$ gauged Nambu-Jona-Lasinio model depending on the gluon fields,
rotated electromagnetic field, diquark condensates, and chemical
potentials,
\begin{equation} \label{effective-action}
\Gamma_{eff}=-\int d^{4}x \{{\frac{1}{4}(f_{\mu
\nu})^{2}+\frac{1}{4}(F_{\mu \nu}^{a})^2}\} + \Gamma_{g}
-\frac{|\Delta|^2\beta V}{4G}+\Gamma_{q},
\end{equation}
where $a=1,...,8$, $G$ is the four-fermion coupling, $\beta$ the
inverse temperature, $V$ the 3-dimensional volume, $f_{\mu
\nu}=\partial_{\mu}A_{\nu}-\partial_{\nu}A_{\mu}$ is the
conventional electromagnetic strength tensor, $\Gamma_{g}$ is the
gauge-fixing action, and $\Gamma_{q}=\frac{1}{2}Tr \ln(
S^{-1}+G_{\mu}^{a}\Gamma^{\mu}_{a}+A_{\mu}\Gamma^{\mu})$ is the
fermion contribution obtained after integrating out the quark fields
in the grand partition function of the neutral $2SC$ phase
\cite{Rischke}. In $\Gamma_{q}$, $S$ is the quasiparticle propagator
in the Nambu-Gor'kov space;
$\Gamma^{\mu}_{a}=diag(g\gamma^{\mu}T_{a},-g\gamma^{\mu}T_{a}^{\intercal})$
is the Nambu-Gor'kov gluon vertex, and
$\Gamma^{\mu}=diag(e\gamma^{\mu}Q,-e\gamma^{\mu}Q)$ is the
Nambu-Gor'kov electromagnetic vertex. In principle, $S$ depends on
the diquark condensate, chemical potentials and possible condensates
of gauge fields.

We assume a weak-coupling approximation ( $\alpha_{s}<1$). Then,
expanding the logarithm up to second order in the gauge fields, we
obtain $\Gamma_{q}=\Gamma_{0}-\frac{1}{2}\int d^{4}x
[G_{\mu}^{a}\Pi^{\mu \nu}_{ab}G_{\nu}^{b}+A_{\mu}\Pi^{\mu
\nu}A_{\nu}]$. All the quantities contributing to $\Gamma_{q}$: the
effective action ($\Gamma_{0}=\frac{1}{2}Tr \ln S^{-1}$), as well as
the polarization operators for gluons ($\Pi^{\mu
\nu}_{ab}=\frac{1}{2}Tr[\Gamma^{\mu}_{a}S\Gamma^{\nu}_{b}S]$) and
photons ($\Pi^{\mu \nu}=\frac{1}{2}Tr[\Gamma^{\mu}S\Gamma^{\nu}S]$)
have been found in the hard-dense-loop approximation of the neutral
2SC phase \cite{Igor}. The absence of linear terms in the expansion
of $\Gamma_{q}$ is due to the zero trace in the color index, which
is a consequence of color neutrality, $\langle J^{c} \rangle =0$
(i.e. zero tadpole), required to preserve the theory non-Abelian
gauge symmetry \cite{Vivian}.

In this derivation we  used t'Hooft gauge to avoid mixing between
the gluons and the fluctuations of the diquark condensate. Details
of the implementation of t'Hooft gauge in the 2SC theory can be
found in \cite{Rischke}. The scalar modes associated to the
fluctuations of the diquark condensate come from the spontaneous
breaking of the color gauge group. Therefore, they are unphysical
and only serve to provide the longitudinal degrees of freedom of the
gluons that acquire a mass due to the symmetry breaking. The
decoupling of the scalar modes from the gluons occurs in the t'Hooft
gauge for arbitrary values of the gauge parameter $\lambda$. Later
on we will choose $\lambda =1$ because it simplifies the
calculations in the gluon sector. As always, physical parameters as
Debye and Meissner masses do not depend on $\lambda$.

\section{Inhomogeneous Gluon Condensate and Induced In-medium Magnetic Field}\label{Inhomogeneous gluon condensate}

To investigate the condensation phenomenon triggered by the
tachyonic modes of the charged gluons, we can restrict our analysis
to the gauge sector of the mean-field effective action
(\ref{effective-action}) that depends on the charged gluon fields
and rotated electromagnetic field. For a static solution, one only
needs the leading contribution of the polarization operators in the
infrared limit ($p_{0}=0, |\overrightarrow{p}|\rightarrow 0$). Under
these conditions, the gauge sector of the effective action can be
written as
\begin{eqnarray}
\label{Eff-Act-2} \Gamma_{eff}^{g}& = & \int d^{4}x
\{-\frac{1}{4}(\widetilde{f}_{\mu
\nu})^{2}-\frac{1}{2}|\widetilde{\Pi}_{\mu}K_{\nu}-\widetilde{\Pi}_{\nu}K_{\mu}|^{2}
\nonumber
 \\
& - & [ m_{M}^{2} \delta_{\mu i} \delta_{\nu
i}+(\mu_{8}-\mu_{3})^{2} \delta_{\mu \nu}+
i\widetilde{q}\widetilde{f}_{\mu \nu}] K_{\mu}K_{\nu}^{\dag}\qquad
\nonumber
 \\
 & + &
\frac{g^2}{2}[(K_{\mu})^{2}(K^{\dag}_{\nu})^{2}-(K_{\mu}K^{\dag}_{\mu})^{2}]
+\frac{1}{\lambda}K^{\dag}_{\mu}\widetilde{\Pi}_{\mu}\widetilde{\Pi}_{\nu}K_{\nu}\}\qquad
\end{eqnarray}
where
$m_{M}^{2}=\frac{2\alpha_{s}\overline{\mu}^{2}}{3\pi}[1-\frac{2\delta\mu^{2}}{\Delta^{2}}]$,
is the Meissner mass with $\overline{\mu}=\mu -\frac{\mu_{e}}{6} +
\frac{\mu_{8}}{3}$ and $\alpha_{s}\equiv\frac{g^{2}}{4\pi}$
\cite{Igor}, the last term in (\ref{Eff-Act-2}) comes from the
$\Gamma_{g}$ term in (\ref{effective-action}), taken in the 't Hooft
gauge with gauge-fixing parameter $\lambda$,
$\widetilde{\Pi}_{\mu}=\partial_{\mu}
-i\widetilde{q}\widetilde{A}_{\mu}$ and $\widetilde{f}_{\mu
\nu}=\partial_{\mu}\widetilde{A}_{\nu}-\partial_{\nu}\widetilde{A}_{\mu}$.
In (\ref{Eff-Act-2}) the Debye mass $m_{D}$ was not included since
it will be no substantial for our derivations. The chemical
potential $\mu_{3}$, although is zero in the gapped phase, should be
in principle taken into account in the analysis of the new phase,
since the K-condensate will break the remaining $SU(2)_C$ symmetry.

At this point let us consider for a moment that we have an external
rotated magnetic field $\widetilde{H}$. In this case the effective
action (\ref{Eff-Act-2}) becomes that of a spin-1 charged field in a
magnetic field (see for instance \cite{emilio}). If we work in the
region where $m^{2}_{M}-\mu^{2}_{8}> 0$, the ground state is stable
($\mu_3=0$ in the stable region). Under these circumstances, we know
\cite{Vortices} that when $q\widetilde{H}\geq
\widetilde{q}\widetilde{H}_{c}= m^{2}_{M}-\mu_{8}^{2}$, the
effective magnetic mass of one of the charged gluon modes becomes
imaginary due to the anomalous magnetic moment term
$i\widetilde{q}\widetilde{f}_{\mu \nu}K_{\mu}K_{\nu}^{\dag}$. This
field-induced instability triggers the formation of a gluon-vortex
state characterized by the antiscreening of the magnetic field.

Now, let us go back to the situation of interest in the present
work, that is, a system with no external magnetic field. As usual in
theories with zero-component gauge-field condensates \cite{Linde},
$\mu_{8}$ gives rise to a tachyonic mass contribution. This
tachyonic contribution creates a region of instability in a density
interval for which $m^{2}_{M}$ is still positive. Let us elaborate
on this point. Using the results of \cite{Igor}, one can easily
realize that when the density is decreasing, coming from values
where $\delta \mu < \Delta/\sqrt{2}$, the different chemical
potentials accordingly change in such a way that the pairing
mismatch $\delta\mu$ increases. Consequently, the Meissner mass
$m^{2}_{M}$ decreases and, at some density value, it crosses the
line of $\mu_{8}$, leading to a tachyonic effective Meissner mass
($m^{2}_{M}-\mu^{2}_{8}<0$) for the range of densities between the
crossing point and the value where $\delta \mu = \delta \mu_{c}$,
even though $m^{2}_{M}>0$ in this interval. Once $\delta \mu \geq
\delta \mu_{c}$, the Meissner mass square $m^{2}_{M}$ becomes
negative itself, so the two terms of the effective Meissner mass
give tachyonic contributions. Although the instability in the region
where $m^{2}_{M}>0$ is a direct consequence of the pairing mismatch,
it formally resembles the instability produced by an external
magnetic field on an otherwise stable color superconductor
\cite{Vortices}. The formal role of the magnetic field is played
here by $\mu_{8}$. Borrowing from the experience gained in the case
with external magnetic field, we would expect that this Meissner
instability will also be removed through the spontaneous generation
of an inhomogeneous gluon condensate $\langle K_{i} \rangle$ capable
to induce a rotated magnetic field, thanks to the anomalous magnetic
moment of the spin-1 charged particles. Having this in mind, we
propose the following ansatz
\begin{eqnarray}  \label{condensate}
\langle K_{\mu} \rangle \equiv \frac{1}{\sqrt{2}}\left(
\begin{array}{cc}
\overline{G}_{\mu}\\
0
\end{array}
\right) \ , \quad \overline{G}_{\mu} \equiv
\overline{G}(x,y)(0,1,-i,0),
\end{eqnarray}
where we took advantage of the $SU(2)_{c}$ symmetry to write the
$\langle K_{i} \rangle$-doublet with only one nonzero component.
Since in this ansatz the inhomogeneity of the gluon condensate is
taken in the $(x,y)$-plane, it follows that the corresponding
induced magnetic field will be aligned in the perpendicular
direction, i.e. along the z-axes, $\langle\widetilde{f}_{12}
\rangle=\widetilde{B}$. The part of the free energy density that
depends on the gauge-field condensates,
$\mathcal{F}_{g}=\mathcal{F}-\mathcal{F}_{n0}$, with
$\mathcal{F}_{n0}=-\Gamma_{0}=\Omega_{0}$ denoting the system
free-energy density in the absence of the gauge-field condensate
($\overline{G}=0, \widetilde{B}=0$), is found, after fixing the
gauge parameter to $\lambda=1$ and using the ansatz
(\ref{condensate}) in (\ref{Eff-Act-2}), to be
\begin{eqnarray}
\label{free-energy} \mathcal{F}_{g} =
\frac{\widetilde{B}^{2}}{2}-2\overline{G}^{\ast}\widetilde{\Pi}^{2}
\overline{G}+2g^{2}|\overline{G}|^{4}\qquad\qquad \nonumber
\\
 -2[2\widetilde{q}\widetilde{B}+(\mu_{3}+\mu_{8})^2+m_{M}^{2}]|\overline{G}|^{2}\qquad
\end{eqnarray}
From the neutrality condition for the $3^{rd}$-color charge it is
found that $\mu_{3}=\mu_{8}$. The fact that $\mu_{3}$ gets a finite
value just after the critical point $m^{2}_{M}-\mu^{2}_{8} = 0$ is
an indication of a first-order phase transition, but since $\mu_{8}$
is parametrically suppressed in the gapped phase by the quark
chemical potential $\mu_{8}\sim \Delta^{2}/\mu$ \cite{Igor}, it will
be a weakly first-order phase transition. Henceforth, we will
consider that $\mu_{3}=\mu_{8}$ in (\ref{free-energy}), and work
close to the transition point $\delta \mu \gtrsim \delta \mu_{c}$
which is the point where $m_{M}^{2}$ continuously changes sign to a
negative value. For very small, negative values of $m_{M}^{2}$, the
gluon condensate and the induced magnetic field should be very small
too, thereby facilitating the calculations.

Minimizing (\ref{free-energy}) with respect to $\overline{G}^{*}$
gives
\begin{equation}
\label{G-Eq} -\widetilde{\Pi}^{2}
\overline{G}-(2\widetilde{q}\widetilde{B}+m_{M}^{2})\overline{G}+2g^{2}|\overline{G}|^{2}\overline{G}=0
\end{equation}
Eq. (\ref{G-Eq}) is a highly non-linear differential equation that
can be exactly solved only by numerical methods. Nevertheless, we
can take advantage of working near the transition point, where we
can manage to find an approximated solution that will lead to a
qualitative understanding of the new condensate phase. With this
aim, and guided by the experience with external fields, where the
solution is always such that the kinetic term
$|\widetilde{\Pi}_{\mu}K_{\nu}-\widetilde{\Pi}_{\nu}K_{\mu}|^{2}$ is
approximately zero near the transition point, we will consider that
when $\delta\mu \simeq \delta\mu_{c}$ our solution will satisfy the
same condition. Hence, we will look for a minimum solution
satisfying
\begin{equation}
\label{G-Eq-2} \widetilde{\Pi}^{2}
\overline{G}+\widetilde{q}\widetilde{B}\overline{G} \simeq 0.
\end{equation}
With the help of (\ref{G-Eq-2}) one can show that the minimum
equation for the induced field $\widetilde{B}$ takes the form
\begin{equation}
\label{B-Eq} 2\widetilde{q} |\overline{G}|^{2}-\widetilde{B}\simeq 0
\end{equation}
The relative sign between the two terms in Eq. (\ref{B-Eq}) implies
that for $|\overline{G}|\neq 0$ a magnetic field $\widetilde{B}$ is
induced. The origin of that possibility can be traced back to the
anomalous magnetic moment term in the action of the charged gluons.
This effect has the same physical root as the paramagnetism found in
Ref. \cite{Vortices}; where contrary to what occurs in conventional
superconductivity, the resultant in-medium field $\widetilde{B}$
becomes stronger than the applied field $\widetilde{H}$ that
triggers the instability. The antiscreening of a magnetic field by
the condensation of charged spin-1 fields has been also found in the
context of the electroweak theory \cite{Olesen}.

Using the minimum equations (\ref{G-Eq}) and (\ref{B-Eq}) in
(\ref{free-energy}), we obtain the condensation free-energy density
\begin{equation}
\label{F-min} \mathcal{\overline{F}}_{g} \simeq
-2(g^2-\widetilde{q}^2)|\overline{G}|^4
\end{equation}
The hierarchy between the strong ($g$) and the electromagnetic
($\widetilde{q}$) couplings implies that
$\mathcal{\overline{F}}_{c}< 0$. Therefore, although the induction
of a magnetic field always costs energy (as can be seen from the
positive first term in (\ref{free-energy})), the field interaction
with the gluon anomalous magnetic moment, produces a sufficiently
large negative contribution to compensate for that increase.
Consequently, as seen from (\ref{F-min}), the net effect of the
proposed condensates is to decrease the system free-energy density.

It follows from Eqs.(\ref{G-Eq})-(\ref{B-Eq}) that near the phase
transition point the inhomogeneity of the condensate solution should
be a small but nonzero correction to a leading constant term
\begin{equation}
\label{Constraint-3} |\overline{G}|^{2}\simeq
\Lambda_{g/\widetilde{q}} m_{M}^{2}/2\widetilde{q}^{2} +\mathcal
{O}(m_{M}^4)f(x,y),
\end{equation}
\begin{equation}
\label{Constraint-2} \widetilde{q}\widetilde{B}\simeq
\Lambda_{g/\widetilde{q}} m_{M}^{2}+\mathcal {O}(m_{M}^4)g(x,y).
\end{equation}
with
$\Lambda_{g/\widetilde{q}}\equiv(g^{2}/\widetilde{q}^{2}-1)^{-1}$.

The explicit form of the inhomogeneity can be found from
(\ref{G-Eq-2}), which can be written in polar coordinates  as
\begin{equation}
\label{Vortex-Eq} [\frac{1}{r}\partial_{r}(r\partial_{r})
+\frac{1}{r^2}\partial_{\theta}^2+\frac{1}{\xi^2}(1-i\partial_{\theta})-\frac{r^{2}}{4\xi^4}]G(r,\theta)=0
\end{equation}
In the above equation we approached the rotated magnetic field by
its leading in (\ref{Constraint-2}), used the symmetric gauge
$\widetilde{A}_{i}=-(\widetilde{B}/2)\epsilon_{ij}x_{j}$, and
introduced the characteristic length $\xi^{2}\equiv
1/\Lambda_{g/\widetilde{q}}m_{M}^{2}$. Using just the leading
contribution of $\widetilde{B}$ in (\ref{Vortex-Eq}) is a consistent
approximation if we simultaneously drop the $\frac{r^{2}}{4\xi^4}$
term and restrict the solution to the domain $r\ll\xi$. Notice that
this domain is indeed a large region because near the critical point
the magnitude of $m_{M}$ is very small, hence $\xi$ becomes very
large. The most symmetric solution of Eq.(\ref{Vortex-Eq}) is the
one that preserves the SO(2) symmetry in the ($x,y$) plane. Hence,
proposing a solution of the form $G(r,\theta)\sim R(r)e^{i\chi}$,
with $\chi$ a constant phase, the equation for $R(r)$ reduces to
$[r\partial_{r}(r\partial_{r}) +\frac{r^{2}}{\xi^2}]R(r)=0$. It is
solved by the Bessel function of the first kind $J_0(r/\xi)$. Then,
the gluon condensate can be written as
$G(r)=(1/\sqrt{2}\widetilde{q}\xi)J_0(r/\xi)\exp i\chi$, which is
consistent with (\ref{Constraint-3}), as in the domain of validity
of this solution ($r\ll \xi$) the series can be approximated by its
first terms. Accordingly, the modulus of the condensate square
satisfies
\begin{equation}
\label{Amplitude}
|\overline{G}|^2\simeq\frac{1}{2\widetilde{q}^2\xi^2}-\frac{r^2}{4\widetilde{q}^2\xi^4}
\end{equation}
The improved solution for $\widetilde{B}$ is found substituting
(\ref{Amplitude}) back into (\ref{B-Eq}). The induced field
$\widetilde{B}$ is homogeneous in the $z$-direction and
inhomogeneous in the $(x,y)$-plane.

As always, the choosing of a particular gauge condition ($\lambda=1$
in this case) is dictated by convenience. As occurs in other gauge
theories, the particular form of the condensate solution
$G(r,\theta)$ and the effective action (or the effective potential
in the case of constant condensates) are gauge-dependent. There is
nothing wrong in that, as none of these quantities is physically
measurable. What is gauge independent is the effective action
evaluated in the minimum solution. Therefore, if the result
(\ref{Amplitude}) is plugged back into (\ref{F-min}), one obtains a
gauge-independent quantity representing the free-energy density of
the ground state.

One may wonder whether this inhomogeneous gluon condensate forms a
vortex state. To answer this question we can compare our results
with the case with external magnetic field \cite{Vortices}. For this
we should notice that the mathematical problem we have just solved
is formally similar to that where the instability is induced by a
weak external field.  This would be the situation when the $2SC$
system approaches the transition point from the stable side (real
effective magnetic mass) and the external magnetic field is slightly
larger than a very small but still positive effective mass square
$\widetilde{H}\gtrsim \widetilde{H}_{c}= m_{M}^{2}-\mu^{2}_{8} \ll
1$. We know that at large $m_{M}^{2}-\mu^{2}_{8} $ the condensate
solution is a crystalline array of vortex-cells with cell's size
$\sim \xi\ll1$. At smaller $m_{M}^{2}-\mu^{2}_{8} $ the lattice
structure should remain, but with a larger separation between cells,
since in this case $\xi\gg1$. However, the use of a linear
approximation to solve the equations in this case only allows to
explore the solution inside an individual cell ($r\ll\xi$). This is
the same limitation we have in the linear approach followed in the
present paper. Therefore, we expect that when the nonlinear
equations will be solved, the vortex arrangement will be explicitly
manifested.

The idea of removing the $2SC$ chromomagnetic instability with the
help of a gluon condensate has been previously considered in Ref.
\cite{miransky}. In that work a homogeneous gluon condensate with no
rotated magnetic field was proposed. Near the transition point the
homogeneous condensate produces a contribution to the condensation
energy density that can be approximated as
$\mathcal{\overline{F}}_{g}\simeq-\frac{\pi}{2\alpha^{3}_{s}}(m^{2}_{M})^{3}/m_{g}^2$,
with $m_{g}^2=4\alpha_{s}^{2} \overline{\mu}^{2}/3\pi$. In this
case, the condensation energy in a circle of radius $c \xi$, with
$c$ being a dimensionless constant satisfying $c<1$, is
${\overline{F}}_{g}\simeq
-\frac{c^{2}\pi^{2}}{2\alpha^{3}_{s}\Lambda_{g/\widetilde{q}}}\frac{m_{M}^4}{m_{g}^2}$.
On the other hand, for the $G-\widetilde{B}$ condensate considered
in the present paper, the corresponding condensation energy is
$\overline{F}_{g}\simeq-c^{2}\pi m_{M}^2/2\widetilde{q}^{2}$, which
is found integrating the free-energy density (\ref{F-min}) evaluated
in the solution (\ref{Amplitude}) in the circle of radius $c\xi$.
Taking into account that near the transition point $m_{M}^{2}\ll
m_{g}^2$ \cite{Igor}, one can see that the inhomogeneous gluon
condensate is energetically favored over the homogeneous one.

We should stress that as $\delta \mu-\delta \mu_{c}$ becomes much
larger, the contribution of $\overline{G}$ and $\widetilde{B}$ in
the quasiparticle propagator $S$ cannot be neglected, thereby
affecting the gap equation (as shown in \cite{MCFL}, a sufficiently
strong magnetic field can qualitatively influence the gap). As a
consequence, the inhomogeneity of the gluon condensate will be
naturally transferred to the diquark condensate. Hence, a LOFF-type
phase may appear as a back reaction of the inhomogeneous gluon
condensate on the gap solution. It is natural to expect that the
additional reduction in free-energy due to the $G-\widetilde{B}$
condensation will make this phase energetically favored over a pure
LOFF one \cite{LOFF}.

We anticipate that a $G-\widetilde{B}$ condensate will likely remove
the chromomagnetic instability in $gCFL$ too. In the gCFL case there
are four gluons with tachyonic masses. Following the results of the
last paper in Ref. \cite{Fukushima}, the four tachyonic gluons are
$A_1$, $A_2$ and two combinations of $A_3$, $A_8$ and $A_\gamma$. A
third combination of $A_3$, $A_8$ and $A_\gamma$ is massless, hence
it represents the rotated electromagnetic field in the gCFL phase.
Notice that, in the unstable gCFL phase the gluons $A_1$ and $A_2$
acquire rotated charge since they couple to the new rotated
electromagnetic field through its $A_3$ component. This implies that
$A_1$ and $A_2$ are analogous to the charged gluons that become
tachyonic in the 2SC case. If $A_1$ and $A_2$ condense in an
inhomogeneous condensate following the same mechanism we found in
the present paper, this condensate could induce a rotated magnetic
field and also give real masses to the two combinations of $A_3$,
$A_8$ and $A_\gamma$ that were tachyonic. Of course, it is possible
too that in addition to the condensation of the charged gluons, the
other two tachyonic combinations also condense in the gCFL case. The
only way to find out which scenario is the most energetically
favored is by doing explicit calculations. However, exploring this
idea will be more challenging, as one will have to deal with a
strongly first-order phase transition.

\section{Astrophysical Connotations}\label{Astrophysical Connotations}

A common characteristic of neutron stars is their strong
magnetization. Their surface magnetic fields range from $B=1.7\times
10^{8} G$ (PSR B1957+20) up to $2.1\times 10^{13} G$ (PSR B0154+61),
with a typical value of $10^{12} G$ \cite{Taylor}. There are
observational discoveries of even strongly magnetized stellar
objects- known as magnetars- with surface magnetic fields of order
$B\sim 10^{14}-10^{16} G$ \cite{magnetars}. In the core of these
compact objects, the field may be considerably larger due to flux
conservation during the core collapse. By applying the scalar virial
theorem it can be shown that the interior field can reach values of
order $B \sim 10^{18}G$ \cite{Lai}.

The observed stellar magnetic fields are supposed to be created by a
magnetohydrodynamic dynamo mechanism that amplifies a seed magnetic
field due to a rapidly rotating protoneutron star. Thus, the
standard explanation of the origin of the magnetar's large magnetic
field implies that the rotation should have a spin period $<3 ms$.
Nevertheless, this mechanism cannot explain all the features of the
supernova remnants surrounding these objects
\cite{magnetar-criticism}.

It has been found \cite{Igor} that when $\delta\mu= \Delta$ in the
neutral 2SC model, the absolute value of the magnetic mass becomes
of order $m_{g}$. This implies that the inhomogeneous gluon
condensate found in our paper could produce a magnetic field of
order of magnitude in the range $\sim 10^{16}-10^{17} G$. As
discussed in Refs. \cite{Phases}, the possibility of generating a
magnetic field of such a large magnitude in the core of a compact
star, without relying on a magneto-hydrodynamics effect, can be an
interesting alternative to address the main criticism
\cite{magnetar-criticism} to the observational conundrum of the
standard magnetar's paradigm \cite{magnetars}. On the other hand, to
have a mechanism that associates the existence of high magnetic
fields to CS at moderate densities can serve to single out the
magnetars as the most probable astronomical objects for the
realization of this high-dense state of matter.

{\bf Acknowledgments:} We are grateful to M. Alford, L. McLerran,
V.~A.~Miransky and I.~A.~ Shovkovy for insightful discussions and
comments. This work has been supported in part by DOE Nuclear Theory
grant DE-FG02-07ER41458.


\begin{thebibliography}{}

\bibitem{Bailin-Love}D.~Bailin, and A.~Love,
Phys. Rep. {\bf 107}, 325 (1984).

\bibitem{Alford} M. Alford, C. Kouvaris, and K. Rajagopal, Phys. Rev. Lett. \textbf{92}, 222001
(2004).

\bibitem{alf-raj-wil-99/537}
M. Alford, K. Rajagopal and F. Wilczek, Phys. Lett. B \textbf{422},
247 (1998); Nucl. Phys. B \textbf{537}, 443 (1999); R. Rapp, T.
Schafer, E. V. Shuryak, and M. Velkovsky, \textit{Phys. Rev. Lett.}
{\bf 81} 53 (1998).

\bibitem{Igor} M. Huang and I. A. Shovkovy, \prd {\bf 70}, 051501 (2004); {\bf 70}, 094030 (2004).

\bibitem{Fukushima}R.~Casalbuoni, et. al.
Phys. Lett. B \textbf{605}, 362 (2005);  \textbf{615}, 297(E)
(2005); M. Alford and Q. H. Wang, J. Phys. G \textbf{31}, 719
(2005); K. Fukushima, Phys. Rev. D \textbf{72}, 074002 (2005).

\bibitem{Schafer}P. F. Bedaque and T. Schafer, Nucl. Phys. A \textbf{697}, 802
(2002); A. Kryjevski and T. Schafer, Phys. Lett. B \textbf{606}, 52
(2005); A. Gerhold and T. Schafer, Phys. Rev. D \textbf{73}, 125022
(2006).

\bibitem{LOFF} M. Alford, J. A. Bowers, and K. Rajagopal, Phys. Rev. D. \textbf{63},
074016 (2001); I. Giannakis and H.C. Ren, Phys. Lett. B
\textbf{611}, 137 (2005); Nucl. Phys. B \textbf{723}, 255 (2005);
R.~Casalbuoni, et.al, Phys. Lett. B \textbf{627}, 89 (2005); M.
Ciminale, G.~Nardulli, M. Ruggieri and R. Gatto, Phys. Lett. B
\textbf{636}, 317 (2006); K. Rajagopal and R. Sharma, Phys. Rev. D
\textbf{74}, 094019 (2006).

\bibitem{miransky} E.~V.~Gorbar, M~Hashimoto, and V.~A.~Miransky, Phys. Lett. B \textbf{632}, 305
(2006); hep-ph/0701211.

\bibitem{Huang}I. Shovkovy and M. Huang, Phys. Lett. B \textbf{564}, 205 (2003); Nucl. Phys. A \textbf{729}, 835 (2003).

\bibitem{Vortices} E.~J.~Ferrer, and V.~de la Incera, Phys.\ Rev.\ Lett.\  {\bf 97}, 122301 (2006); J. Phys. A: Math. Theor. {\bf 40}, 6913 (2007).

\bibitem{zero-mode}V. V. Skalozub, \textit{Sov. J. Nucl. Phys.} \textbf{28},
113 (1978); N. K. Nielsen and P. Olesen, \textit{Nucl. Phys. B}
\textbf{144}, 376 (1978); V. V. Skalozub, \textit{Sov. J. Nucl.
Phys.} \textbf{43}, 665 (1986); \textbf{45}, 1058 (1987); S. Ferrara
and M. Porrati, \textit{Mod. Phys. Lett. A} \textbf{8}, 2497 (1993);
E. J. Ferrer and V. de la Incera, \textit{Int. Jour. of Mod. Phys.
A} \textbf{11}, 3875 (1996).

\bibitem{Olesen}J. Ambjorn and P. Olesen, Nucl. Phys. B \textbf{315},
606 (1989); Phys. Lett. B \textbf{218}, 67 (1989).

\bibitem{Cristina}D. Litim and C. Manuel, Phys. Rev. D \textbf{64}, 094013 (2001).

\bibitem{Rischke}D. H. Rischke, and I. A. Shovkovy, Phys. Rev. D \textbf{66}, 054019 (2002).

\bibitem{Vivian} E.~J.~Ferrer, V.~de la Incera, and A.E. Shabad, Nuovo Cim. A {\bf 98},
245 (1987).

\bibitem{emilio}
E. Elizalde and E. J. Ferrer, and V. de la Incera, Ann. of Phys.
{\bf 295}, 33 (2002); \prd {\bf 70}, 043012 (2004).

\bibitem{Linde}  A. D. Linde, Phys. Lett. B {\bf 86} 39 (1979); E. J. Ferrer, V. de la Incera and A. E. Shabad, Phys.
Lett. B {\bf 185} 407 (1987); Nucl. Phys. B {\bf 309} 120 (1988).

\bibitem{MCFL}
E.~J.~Ferrer, V.~de la Incera and C.~Manuel,
Phys.\ Rev.\ Lett.\  {\bf 95}, 152002 (2005); Nucl. Phys. B
\textbf{747}, 88 (2006).

\bibitem{Taylor} I. Fushiki, E. H. Gudmundsson and C. J. Pethick, \textit{Astrophys.
J.} \textbf{342}, 958 (1989); T. A. Mihara, et al., \textit{Nature
(London)} \textbf{346}, 250 (1990); G. Chanmugam, \textit{Ann. Rev.
Astron. Astrophys.} \textbf{30}, 143 (1992); J. H. Taylor, R. N.
Manchester and A. G. Lyne, \emph{Astrophys. J. S.} {\bf 88}, 529
(1993); P. P. Kronberg, \textit{Rep. Prog. Phys.} \textbf{57}, 325
(1994); D. Lai \textit{Rev. Mod. Phys.} {\bf 73}, 629 (2001); D.
Grasso and H. R. Rubinstein \textit{Phys. Rep.} \textbf{348}, 163
(2001).

\bibitem{magnetars}
C.~Thompson and R.~C.~Duncan, \textit{Astrophys. J.} {\bf 392}, L9
(1992); {\bf 473}, 322 (1996); S. Kulkarni and D. Frail,
\textit{Nature} {\bf 365}, 33 (1993); T. Murakami et al.,
\textit{Nature} {\bf 368}, 127 (1994); Ibrahim et al.,
\textit{Astrophys. J.} {\bf 609}, L21 (2004).

\bibitem{Lai} L.~Dong and S.~L.~Shapiro \textit{Astrophys. J.} {\bf 383}, 745 (1991).

\bibitem{magnetar-criticism}
J. Vink and L. Kuiper, \textit{Mon.\ Not.\ Roy.\ Astron.\ Soc.\
Lett.}  {\bf 370} (2006) L14; R-X Xu, astro-ph/0611608.

\bibitem{Phases} E.~J.~Ferrer, and V.~de la Incera, \prd {\bf 76}, 045011 (2007) (nucl-th/0703034); E. J. Ferrer, arXiv:0705.2403 [hep-ph] (To
appear in the AIP Proceedings of the VII Latin American Symposium on
Nuclear Physics and Applications.)






\end{thebibliography}
\end{document}